\documentclass[aps,prd,showpacs,showkeys,floatfix,nofootinbib,twocolumn,
10pt]{revtex4-1}
\usepackage{tabularx}
\usepackage{natbib}
\usepackage{ulem,mwe}
\usepackage{color}
\usepackage{graphicx,verbatim}
\usepackage{epsfig}
\usepackage{bm}
\usepackage{amsthm}
\usepackage{amsfonts}
\usepackage{float}
\usepackage{amsmath,amssymb}
\usepackage[caption = false]{subfig}
\usepackage[colorlinks=true, citecolor=blue, urlcolor=blue ]{hyperref}
\usepackage{cleveref}
\usepackage{xcolor}
\begin{document}

\title{Systematics of chemical freeze-out parameters in heavy-ion
collision experiments}

\author{Sumana Bhattacharyya}
\email{response2sumana@jcbose.ac.in}
\author{Deeptak Biswas}
\email{deeptak@jcbose.ac.in}
\author{Sanjay K. Ghosh}
\email{sanjay@jcbose.ac.in}
\author{Rajarshi Ray}
\email{rajarshi@jcbose.ac.in}
\author{Pracheta Singha}
\email{pracheta@jcbose.ac.in}
\affiliation{
Department of Physics,
\\ \& \\ 
	Center for Astroparticle Physics \& Space Science,\\
Bose Institute, EN-80, Sector-5, Bidhan Nagar, Kolkata-700091, India 
}

\begin{abstract}
We discuss systematic uncertainties in the chemical freeze-out
parameters from the $\chi^2$ analysis of hadron multiplicity ratios in
the heavy-ion collision experiments. The systematics due to the choice
of specific hadron ratios are found to lie within the experimental
uncertainties. The variations obtained by removing the usual constraints
on the conserved charges show similar behavior. The net charge to net
baryon ratios in such unconstrained systems are commensurate with the
expected value obtained from the protons and neutrons of
the colliding nuclei up to the center of mass energies $\sim 40$ GeV.
Beyond that the uncertainties in this ratio gradually increases,
possibly indicating the reduction in baryon stopping.

\end{abstract}

\keywords{ Heavy Ion collision, Chemical freeze-out, Hadron Resonance
Gas model} \maketitle

\section{\label{sec:Intro} Introduction}

Experimental data from the relativistic heavy-ion collisions indicate
that a considerable fraction of energies of the colliding nuclei are
deposited in the region of interaction to form a hot and dense fireball.
The resulting system may thermalize rapidly to form the exotic state of
thermally equilibrated quark gluon matter~\cite{PhysRevD.22.2793,
PhysRevLett.48.1066, Koch:1986ud}.  Pressure gradients from the central
to the peripheral region of the system would lead to expansion and
cooling. Thereafter the system would most likely make a transition from
the partonic phase to the hadronic phase in thermal and chemical
equilibrium~\cite{Hagedorn:1965st, Hagedorn:1980kb}.  With further
expansion, and increase in inter-particle separation, the chemical
composition of the hadrons would freeze out. The characteristics of the
freeze-out surface depends on the nature of interaction or
scattering~\cite{Heinz:1998st}. 

Several statistical models of hadronic system has been formulated in
last few years, based on the seminal work by Fermi~\cite{Fermi:1950jd},
followed by Pomeranchuk~\cite{Pomeranchuk:1951ey} and
Landau~\cite{Landau:1953gs,Belenkij:1956cd} in subsequent years.
Hagedorn recognized that the existence of numerous hadronic resonances
is a characterization of hadron's strong
interactions~\cite{Hagedorn:1965st}. These simple statistical approaches
are successful with surprising accuracy in explaining the yield of
experimentally observed hadrons produced in collisions of elementary
particles and as well as heavy-ions~\cite{PhysRevLett.81.5284,
Rischke:1991ke, Cleymans:1992jz, BraunMunzinger:1994xr, Cleymans:1996cd,
Yen:1997rv, Heinz:1998st, BraunMunzinger:1999qy, Cleymans:1999st,
BraunMunzinger:2001ip, Becattini:2003wp, BraunMunzinger:2003zd,
Karsch:2003zq, Tawfik:2004sw, Becattini:2005xt, Andronic:2005yp,
Andronic:2008gu, Manninen:2008mg, Tiwari:2011km, Begun:2012rf,
Andronic:2012ut, Fu:2013gga, Tawfik:2013eua, Garg:2013ata,
Bhattacharyya:2013oya, Albright:2014gva, Kadam:2015xsa, Kadam:2015fza,
Kadam:2015dda, Albright:2015uua, Bhattacharyya:2015zka,
Bhattacharyya:2015pra, Begun:2016cva, Bhattacharyya:2017gwt,
Andronic:2017pug, Ghosh:2018nqi, Dash:2018can}. The underlying
consideration in these analyses are that, even when the hadronic
interactions cease at some point of evolution of the fireball, the
hadrons populate the phase space according to statistical
distribution~\cite{BraunMunzinger:2003zd}.

Assuming the chemical equilibration of the full hadronic spectra, the
thermodynamic freeze-out parameters are obtained from the statistical
model calculations by performing a $\chi^2$ fit with the available
experimental multiplicity data~\cite{Andronic:2005yp, Andronic:2008gu,
Andronic:2012ut, Alba:2014eba, Chatterjee:2015fua, Manninen:2008mg,
Chatterjee:2017yhp, Adak:2016jtk, Cleymans:2004hj,
Cleymans:2005xv,Chatterjee:2013yga}. One of the most important issues
has been the discussions of resulting enhancement of multi-strange
hadrons~\cite{PhysRevLett.43.1292}, as well as the  saturation of
strangeness~\cite{PhysRevLett.48.1066}. However possible scenarios of
departure of strangeness chemical equilibrium introduced through an
additional parameter $-$ the {\it strangeness suppression factor} $
\gamma_s$, have been discussed~\cite{Koch:1986ud, Adamczyk:2017iwn,
Andronic:2005yp, Becattini:2003wp, Becattini:2005xt, Manninen:2008mg,
Cleymans:2005xv}.  More recently the possibility of mass dependent or
flavor dependent freeze-out pictures have emerged (see
e.g.~\cite{Chatterjee:2015fua, Chatterjee:2013yga}). 
The studies in this direction has reached the state-of-the-art with
several sophisticated codes like THERMUS~\cite{Wheaton:2004qb},
SHARE~\cite{Torrieri:2004zz}, THERMINATOR~\cite{Kisiel:2005hn} publicly
available for studying the freeze-out characteristics.

Here we revisit the analysis of the chemical freeze-out parameters from
the $\chi^2$ fit of experimental multiplicity data ratios using the
Hadron Resonance Gas (HRG) model for a different purpose. It has been
generally found that equilibration parameters are biased to the chosen
particle ratios. For example in Ref.~\cite{Andronic:2005yp} a particular
set of independent hadron ratios are used with minimum repetition of
any given hadronic species to reduce systematic bias. We shall instead
consider a variety of possible sets of hadron yield ratios to obtain the
variations of the estimated freeze-out parameters, and the resulting
variations in the predicted yield ratios. This would give us a
quantitative estimate of the systematic uncertainty in the analysis
method due to the choice of the hadron yield ratios.

Another possible source of systematic uncertainty is the choice of
constraints to be satisfied by the system. The most popular choice is to
consider a fixed ratio of net electric charge to net baryon number
depending on the colliding nuclei, along with the setting of net
strangeness to be zero. Usually the available hadronic yield data is
obtained within a limited central rapidity bin. Though the constraints
should definitely be satisfied globally, there is no imperative reasons
to enforce these conditions in a given rapidity bin. For example, at
very high center of mass energy of collisions ($\sqrt{s}$) one expects
baryon stopping in the central rapidity region to be negligible. The
system may simply be composed of the secondary particles produced, and
would have both the net charge and net baryon number to vanish. It is
therefore natural to study the system without considering any
constraints at all. Here we shall perform this analysis and check the
conditions on the net charges when the constraints are not used.

The paper is organized as follows. A brief description of HRG model is
given in section \ref{sec:HRG}.  In section \ref{sec:appl} $-$
\ref{sec:result} we have described our model and its results followed by
discussion. Our conclusions are presented in section
\ref{sec:conclusion}.

\section{\label{sec:HRG}Hadron resonance gas model}

For strong interactions, chemical equilibrium means the equilibration of
the conserved charges baryon number ($B$), electric charge ($Q$) and
strangeness ($S$). Thus, the equilibrium thermodynamic parameters are
the temperature $T$ and the three chemical potentials $\mu_B$, $\mu_Q$
and $\mu_S$, corresponding to the three conserved charges respectively.
At chemical freeze-out, the HRG model may provide a good description of
the system in thermodynamic equilibrium~\cite{BraunMunzinger:1994xr,
Cleymans:1996cd, BraunMunzinger:1999qy, Cleymans:1999st,
BraunMunzinger:2001ip, Becattini:2005xt, Andronic:2005yp,
Andronic:2008gu}. The partition function of HRG in the grand canonical
ensemble is given as,

\begin {equation}
\ln Z^{ideal}=\sum_i \ln Z_i^{ideal},
\end{equation}

\noindent
where sum runs over all the hadrons and resonances. Here thermodynamic
potential for $i$'th species is given as,

\begin{equation}
\ln Z_i^{ideal}=\pm \frac{Vg_i}{(2\pi)^3}\int d^3p
\ln[1\pm\exp(-(E_i-\mu_i)/T)],
\end{equation}

\noindent
where the upper sign is for fermions and lower is for bosons, and $V$ is
the volume of the system. Here $g_i$, $E_i$ and $m_i$ are respectively
the degeneracy factor, energy and mass of $i^{th}$ hadron.  The chemical
potential of the $i^{th}$ hadron may be expressed as,
$\mu_i=B_i\mu_B+S_i\mu_S+Q_i\mu_Q$, with $B_i$, $S_i$ and $Q_i$ denoting
its baryon number, strangeness and electric charge.

\section{\label{sec:appl}Application to Freeze-out}

From the partition function, the number density $n_i$ is obtained as,

\begin{equation}
 n_i =\frac{T}{V} \left(\frac{\partial \ln Z_i}
       {\partial\mu_i}\right)_{V,T} 
 =\frac{g_i}{{(2\pi)}^3} \int\frac{d^3p} {\exp[(E_i-\mu_i)/T]\pm1}.
\end{equation}

\noindent
One may relate this number density for the $i$'th detected hadron to the
corresponding rapidity density as~\cite{Manninen:2008mg},

\begin{equation}
\frac{dN_i}{dy}\Biggr|_{Det} \simeq 
{\frac{dV}{dy}}n_i^{Tot}\Biggr|_{Det}
\end{equation}
\noindent
where the subscript $Det$ denotes the detected hadrons. Here,

\begin{eqnarray}
&n_i^{Tot}& ~=~ n_i(T,\mu_B,\mu_Q,\mu_S)~ + 
\nonumber \\
&\sum_j& n_j(T,\mu_B,\mu_Q,\mu_S) \times Branch~Ratio (j
\rightarrow i)
\end{eqnarray}
\noindent
where the summation is over the unstable resonances $j$ that decay to
the $i^{th}$ hadron. To remove systematics due to the volume factor the
ratios of yields are usually employed. Thus, the expected equations to
be satisfied are,

\begin{eqnarray}
R_{\alpha}^{exp} = \frac{\frac{dN_{i}}{dY}}{\frac{dN_{j}}{dY}} &\simeq&
\frac{n_{i}(T,\mu_B,\mu_Q,\mu_S)}{n_{j}(T,\mu_B,\mu_Q,\mu_S)} =
R_{\alpha}^{therm}
\end{eqnarray}

\noindent
where, $R_{\alpha}^{exp}$ is the ratio of hadron yields obtained from
experiments and $R_{\alpha}^{therm}$ is the ratio of the number
densities from the model calculations. However individual hadron numbers
are not conserved in strong interactions. Rather an equilibrium of the
hadrons in terms of the conserved charges $B$, $Q$ and $S$ are sought.
Therefore a $\chi^2$ fitting is performed with as many such ratios as
permissible. 

It is a common approach to first fix two of the parameters, say, $\mu_Q$
and $\mu_S$ using the constraint relations \cite{Alba:2014eba},  

\begin{equation}
\label{eq:nbq}
\frac{\sum_i n_i (T, \mu_B, \mu_S, \mu_Q) Q_i}{ \sum_i n_i (T, \mu_B, 
\mu_S,\mu_Q) B_i} = constant,
\end{equation}
and
\begin{equation}
\label{eq:ns}
\sum_i n_i (T, \mu_B, \mu_S, \mu_Q) S_i=0.
\end{equation}

\noindent
The value of the ratio of net baryon number to net charge depends on the
physical system. For example, in Au + Au collisions, $constant =
N_p/(N_p + N_n) \simeq 0.4$, with $N_p$ and $N_n$ denoting the number of
protons and neutrons in the colliding nuclei.  The problem is now
reduced to obtaining the best fit value of the other two parameters by
minimizing the $\chi^{2}$, which is defined as,

\begin{equation}
\label{eq:chisqr}
\chi^{2}=\sum_{\alpha} \frac{(R_{\alpha}^{exp}-R_{\alpha}^{therm})^{2}}
{(\sigma^{exp}_{\alpha})^2},
\end{equation}

\noindent
Here $\sigma^{exp}_{\alpha}$ is the experimental uncertainty in the
ratio $R_{\alpha}^{exp}$.  This kind of a setup, and its various
improved versions in terms of the modifications of the HRG model, has
successfully described hadron yields from AGS to LHC energies
~\cite{BraunMunzinger:1994xr, Cleymans:1996cd, Yen:1997rv,
BraunMunzinger:1999qy, Cleymans:1999st, BraunMunzinger:2001ip,
BraunMunzinger:2003zd, Becattini:2003wp, Cleymans:2004hj,
Letessier:2005qe, Becattini:2005xt, Cleymans:2005xv, Andronic:2005yp,
Manninen:2008mg, Andronic:2008gu, Tiwari:2011km, Chatterjee:2013yga,
Alba:2014eba, Chatterjee:2015fua, Adak:2016jtk}. 

\section{\label{sec:syst} Systematic Uncertainties}

There are many possible sources of systematic uncertainties that enter
in the analysis. The systematic uncertainties in the experimental data
are expected to be reduced when considering the ratios of hadron yields.
For example, in Ref.~\cite{Andronic:2005yp, Adamczyk:2017iwn} more
particle-antiparticle ratios are used than the cross ratios, as they may
achieve some cancellation of the inherent systematic errors of
experimental measurements. 

\subsection{Systematics due to choice of hadron ratios}

However the use of the hadron ratios introduces further systematics into
the picture. Therefore different authors consider different criteria for
choosing the set of hadron ratios. For example
Ref.~\cite{Andronic:2005yp} claimed that whenever a given experimental
yield is used several times in the ratios there may be some correlation
uncertainties in particle ratios. An estimate of such uncertainties was
found to be less than 5 percent~Ref.~\cite{Adamczyk:2017iwn}.  On the
other hand, choosing all possible hadron yield ratios do not serve the
purpose either, as there will be relevant correlations between different
ratios ~\cite{Andronic:2005yp,Becattini:2007wt}. The alternate
possibility is to choose $(N-1)$ number of statistically independent
ratios of $N$ hadron yields.  But it has been argued that this
procedure may lead to the information loss \cite{Becattini:2007wt,
Manninen:2008mg}. 

Here we propose to quantify the systematic uncertainty due to the
consideration of different sets of statistically independent hadron
yield ratios. With the $N$ experimentally measured hadron yields one can
form $^NC_2=N(N-1)/2$ yield ratios (interchanging numerator and
denominator does not give any new information). Any $(N-1)$ of these
ratios are statistically independent as long as each of the $N$ hadron
yields appear at least once. For example, if there are three hadrons
$\pi^+$, $\pi^-$ and $p$, there are in all $^3C_2=3$ possible ratios
that can be formed. Out of these 3 ratios only $(3-1)=2$ are
independent, because the third ratio can always be written as a
combination of the 2 independent ratios. However any 2 of these three
ratios can be chosen to be independent, and there are 3 different ways
to choose them.  For higher number of hadrons, all the sets of $(N-1)$
ratios may not necessarily be independent. Still the number of
independent sets would grow very large with the increase in number of
hadron yields.  In general, for $N$ given hadron yields there are
$N(N-1)/2$ possible ratios, from which $(N-1)$ ratios may be chosen in
$^{N(N-1)/2}C_{N-1}$ ways. 

One can obtain the freeze-out parameters for each of these sets of
independent ratios. The variation of the parameters with the choice of
these different sets would then be an useful measure of the concerned
systematic uncertainty.

For the constrained system, we shall use Eq.~\ref{eq:nbq} and
Eq.~\ref{eq:ns} for fixing $\mu_B$ and $\mu_S$, and fit $T$ and $\mu_Q$
from the minimizing conditions on $\chi^2$ given as,

\begin{equation}
\label{eq:difwtemp}
\frac{\partial\chi^{2}}{\partial T}=0,
\end{equation}
\begin{equation}
\label{eq:difwmuq}
\frac{\partial \chi^{2}}{\partial \mu_Q}=0
\end{equation}
\noindent
We shall perform the $\chi^2$ analysis for each of the chosen sets of
ratios and obtain the variation of the resulting freeze-out parameters.
Finally we shall obtain the corresponding variations of the predicted
hadron yield ratios.

\subsection{Relaxing the net charge constraints}

The other possible systematics is the use of the constraint equations
Eq.~\ref{eq:nbq} and Eq.~\ref{eq:ns} respectively. The constraints
essentially ensure that the overall net baryon to net charge ratio as
well as the vanishing of net strangeness is preserved even locally for
the small rapidity window of the experimental data. However such
restrictions are not governed by any physical laws. In fact for very
large $\sqrt{s}$, say for LHC energies, baryon stopping is expected to
be negligible.  In that case both net baryon number and net charge may
vanish and their ratio may be indeterminate. Therefore one can try to
obtain the freeze-out parameters by varying the constraints. For example
in Ref.~\cite{Adamczyk:2017iwn} variations in the systematics were
obtained by choosing several conditions like ($i$) $\mu_Q=0$, ($ii$)
$\mu_Q$ constrained by net baryon to net charge ratio, and ($iii$)
$\mu_Q$ constrained to that ratio along with vanishing net strangeness.
Similarly there has been studies~\cite{Bravina:1999dh} that discuss
possibilities of net strangeness being non-zero in these regions.  The
most drastic change would be to remove the constraints altogether, which
we shall try here. The thermodynamic parameters will be obtained from
the four $\chi^2$ minimization equations,

\begin{equation}
\label{eq:difwt}
\frac{\partial \chi^{2}}{\partial T}=0,
\end{equation}
\begin{equation}
\label{eq:difwmu}
\frac{\partial \chi^{2}}{\partial \mu_x}=0, \quad \mathrm{where},
 x=B,Q,S.
\end{equation}

\noindent
Here also we shall study the systematics due to the choice of hadron
yield ratios as discussed in the last subsection.  Further it would be
interesting to study the systematics of the net baryon to net charge
ratio and the net strangeness with varying $\sqrt{s}$.

\section{Data Analysis}
We have used 
AGS~\cite{Ahle:1999uy, Ahle:2000wq, Klay:2003zf,
Klay:2001tf, Back:2001ai, Blume:2011sb, Back:2000ru, Barrette:1999ry, 
Back:2003rw}, 
SPS~\cite{Alt:2007aa, Alt:2005gr, Afanasiev:2002mx, Afanasev:2000uu,
Bearden:2002ib, Anticic:2003ux, Antinori:2004ee, Antinori:2006ij,
Alt:2008qm, Alt:2008iv, Anticic:2003ux}, RHIC~\cite{Kumar:2012fb,
Das:2012yq, Adler:2002uv, Adams:2003fy, Zhu:2012ph, Zhao:2014mva,
Kumar:2014tca, Das:2014kja, Abelev:2008ab, Aggarwal:2010ig,
Abelev:2008aa, Adcox:2002au, Adams:2003fy, Adler:2002xv, Adams:2006ke,
Adams:2004ux, Kumar:2012np, Adams:2006ke} and LHC~\cite{Abelev:2012wca,
Abelev:2013xaa, ABELEV:2013zaa, Abelev:2013vea} data for our analysis.
The data for STAR BES is obtained from Ref~\cite{Chatterjee:2015fua,
Nasim:2015gua, Adamczyk:2017iwn}. The data at mid-rapidity for the most
central collisions are considered. All hadrons with mass up to 2 GeV
are included for the HRG spectrum. The masses and branching ratios used
are as given in ~\cite{Wheaton:2004qb, Tanabashi:2018oca}. Experimental
yields are available for only a few hadrons at various collision
energies.  Also the data for all the hadronic yields at one collision
energy may not be available in another. For example, the LHC data set
does not have the $\bar{\Lambda}$, and we assumed it to be same as that
reported for $ {\Lambda}$. Similarly, we did not use $\Omega$ data for
any parametrization, as the individual yields of $\Omega^+$ and
$\Omega^- $ are not available for most of the $\sqrt{s}$. Data for
$\phi$ (1019.46 MeV) was not used in the fitting as it is already
included in the analysis through its strong decay channel to kaon.  So
the yields of the hadrons used to obtain the freeze-out parameters are,
$\pi^\pm$ (139.57 MeV), $k^{\pm}$ (493.68 MeV), $p$,$\bar{p}$ (938.27
MeV), $\Lambda$,$\bar{\Lambda}$ (1115.68 MeV), $\Xi^\mp$ (1321.71 MeV).
For the AGS at $\sqrt{s}=4.85$ GeV, $\Xi$ data was not available.
For the lower AGS energies we used $\pi^\pm$, $k^{\pm}$, $p$
and $\Lambda$ data.

\noindent
\begin{table}[!htb]
\begin{tabular}{|c|ccccccccc|}
\hline
& \\[-2 ex]
Set 1 & $\frac{\pi^{-}}{\pi^{+}}$, & $\frac{k^{+}}{\pi^{+}}$, &
$\frac{k^{-}}{\pi^{+}}$, & $\frac{p}{\pi^{+}}$, &
$\frac{\bar{p}}{\pi^{+}}$, &
$\frac{\Lambda}{\pi^{+}}$, & $\frac{\bar{\Lambda}}{\pi^{+}}$, &
$\frac{\Xi^{-}}{\pi^{+}}$, & $\frac{\Xi^{+}}{\pi^{+}}$\\[1 ex]
\hline
\hline
& \\[-2 ex]
Set 2 & $\frac{\pi^{-}}{\pi^{+}}$, & $\frac{k^{+}}{\pi^{-}}$, &
$\frac{k^{-}}{\pi^{-}}$, & $\frac{p}{\pi^{-}}$, &
$\frac{\bar{p}}{\pi^{-}}$, &
$\frac{\Lambda}{\pi^{-}}$, & $\frac{\bar{\Lambda}}{\pi^{-}}$, &
$\frac{\Xi^{-}}{\pi^{-}}$, & $\frac{\Xi^{+}}{\pi^{-}}$\\[1 ex]
\hline
\hline
& \\[-2 ex]
Set 3 & $\frac{\pi^{-}}{\pi^{+}}$, & $\frac{k^{+}}{\pi^{-}}$, &
$\frac{k^{-}}{k^{+}}$, & $\frac{p}{k^{+}}$, & $\frac{\bar{p}}{k^{+}}$, &
$\frac{\Lambda}{k^{+}}$, & $\frac{\bar{\Lambda}}{k^{+}}$, &
$\frac{\Xi^{-}}{k^{+}}$, & $\frac{\Xi^{+}}{k^{+}}$\\[1 ex]
\hline
\hline
& \\[-2 ex]
Set 4 & $\frac{\pi^{-}}{\pi^{+}}$, & $\frac{k^{+}}{\pi^{-}}$, &
$\frac{k^{-}}{k^{+}}$, & $\frac{p}{k^{-}}$, & $\frac{\bar{p}}{k^{-}}$, &
$\frac{\Lambda}{k^{-}}$, & $\frac{\bar{\Lambda}}{k^{-}}$, & 
$\frac{\Xi^{-}}{k^{-}}$, & $\frac{\Xi^{+}}{k^{-}}$\\[1 ex]
\hline
\hline
& \\[-2 ex]
Set 5 & $\frac{\pi^{-}}{\pi^{+}}$, & $\frac{k^{+}}{\pi^{-}}$, &
$\frac{k^{-}}{k^{+}}$, & $\frac{p}{k^{-}}$, & $\frac{\bar{p}}{p}$, &
$\frac{\Lambda}{p}$, & $\frac{\bar{\Lambda}}{p}$, & $\frac{\Xi^{-}}{p}$,
&
$\frac{\Xi^{+}}{p}$\\[1 ex]
\hline
\hline
& \\[-2 ex]
Set 6 & $\frac{\pi^{-}}{\pi^{+}}$, & $\frac{k^{+}}{\pi^{-}}$, &
$\frac{k^{-}}{k^{+}}$, & $\frac{p}{k^{-}}$, & $\frac{\bar{p}}{p}$, &
$\frac{\Lambda}{\bar{p}}$, & $\frac{\bar{\Lambda}}{\bar{p}}$, & 
$\frac{\Xi^{-}}{\bar{p}}$, & $\frac{\Xi^{+}}{\bar{p}}$\\[1 ex]
\hline
\hline
& \\[-2 ex]
Set 7 & $\frac{\pi^{-}}{\pi^{+}}$, & $\frac{k^{+}}{\pi^{-}}$, &
$\frac{k^{-}}{k^{+}}$, & $\frac{p}{k^{-}}$, & $\frac{\bar{p}}{p}$, &
$\frac{\Lambda}{\bar{p}}$, & $\frac{\bar{\Lambda}}{\Lambda}$, & 
$\frac{\Xi^{-}}{\Lambda}$, & $\frac{\Xi^{+}}{\Lambda}$\\[1 ex]
\hline
\hline
& \\[-2 ex]
Set 8 & $\frac{\pi^{-}}{\pi^{+}}$, & $\frac{k^{+}}{\pi^{-}}$, &
$\frac{k^{-}}{k^{+}}$, & $\frac{p}{k^{-}}$, & $\frac{\bar{p}}{p}$, &
$\frac{\Lambda}{\bar{p}}$, & $\frac{\bar{\Lambda}}{\Lambda}$, & 
$\frac{\Xi^{-}}{\bar{\Lambda}}$, & $\frac{\Xi^{+}}{\bar{\Lambda}}$\\[1 ex]
\hline
\hline
& \\[-2 ex]
Set 9 & $\frac{\pi^{-}}{\pi^{+}}$, & $\frac{k^{+}}{\pi^{-}}$, &
$\frac{k^{-}}{k^{+}}$, & $\frac{p}{k^{-}}$, & $\frac{\bar{p}}{p}$, &
$\frac{\Lambda}{\bar{p}}$, & $\frac{\bar{\Lambda}}{\Lambda}$, & 
$\frac{\Xi^{-}}{\bar{\Lambda}}$, & $\frac{\Xi^{+}}{\Xi^{-}}$\\[1 ex]
\hline
\end{tabular}
\caption{Set of sets of hadron yield ratios for AGS ($\sqrt{s}=4.85$
GeV), SPS, RHIC and LHC data.}
\label{tb.sets}
\end{table}

For the ten hadron yields in SPS, RHIC, and LHC data, one can construct
a set of nine independent ratios as discussed earlier. The ratios in
each set will be independent if each of the ten hadron yields appear at
least once in the nine ratios. Therefore not all choices of nine ratios
will contain independent ratios. There is a huge multiplicity of such
suitable sets.  We selected a sample of only nine such sets as given in
table~\ref{tb.sets}, to make an estimate of the systematic uncertainty
due to the choice of sets.  These sets were chosen such that each of the
$^{10}C_2=45$ ratios appear in at least one set. However some of the
ratios like $\pi^{-}/\pi^{+}$ appear in all of them and some others
appear less.  Even some of the ratios like $k^{+}/\pi^{+}$ and
$\Xi^{+}/\Xi^{-}$ appear only once.

\noindent
\begin{table}[!htb]
\begin{tabular}{|c|c|}
\hline
& \\[-2 ex]
Set 1 & $\frac{\pi^{-}}{\pi^{+}}$, $\frac{k^{+}}{\pi^{+}}$,
$\frac{k^{-}}{\pi^{+}}$, $\frac{p}{\pi^{+}}$
$\frac{\Lambda}{\pi^{+}}$\\[1 ex]
\hline
\hline
& \\[-2 ex]
Set 2 & $\frac{\pi^{-}}{\pi^{+}}$, $\frac{k^{+}}{\pi^{-}}$,
$\frac{k^{-}}{\pi^{-}}$, $\frac{p}{\pi^{-}}$,
$\frac{\Lambda}{\pi^{-}}$\\[1 ex]
\hline
\hline
& \\[-2 ex]
Set 3 & $\frac{k^{-}}{\pi^{+}}$, $\frac{k^{-}}{\pi^{-}}$,
$\frac{k^{-}}{k^{+}}$, $\frac{p}{k^{-}}$,
$\frac{\Lambda}{k^{-}}$\\[1 ex]
\hline
\hline
& \\[-2 ex]
Set 4 & $\frac{p}{\pi^{+}}$, $\frac{p}{\pi^{-}}$,
$\frac{p}{k^{+}}$, $\frac{p}{k^{-}}$,
$\frac{\Lambda}{p}$\\[1 ex]
\hline
\hline
& \\[-2 ex]
Set 5 & $\frac{\Lambda}{\pi^{+}}$, $\frac{\Lambda}{\pi^{-}}$,
$\frac{\Lambda}{k^{+}}$, $\frac{\Lambda}{k^{-}}$,
$\frac{\Lambda}{p}$\\[1 ex]
\hline
\end{tabular}
\caption{Set of sets of hadron yield ratios for lower AGS energies}
\label{tb.sets2}
\end{table}

For the eight hadron yields for the AGS data at $\sqrt{s}=4.85$ GeV we
have used the seven yield ratios excluding those involving $\Xi$, from
the sets 1-7 of table~\ref{tb.sets}. For the six hadron yields for the
lower AGS energies, we used the set of ratios given in
table~\ref{tb.sets2}.

We shall extract freeze-out parameters for each of the given sets. The
variation of the parameters from these sets will be our estimate for the
systematic uncertainties. The central values would be obtained from the
weighted mean. Investigating the effects of choosing few more such sets
may be attempted in future, but we note that the actual number of all
such independent sets is prohibitively large.

All these considerations will remain same when we study the freeze-out
characteristics by removing the constraints on the net charges. Our
numerical code solves the equations Eq.~[\ref{eq:nbq} $-$
\ref{eq:difwmuq}] for the case including the constraints, and the
equations Eq.~[\ref{eq:difwt}$-$~\ref{eq:difwmu}] for the case without
the constraints using the Broyden's method with a convergence criteria
of $10^{-6}$ or better.

\section{\label{sec:result} Results and discussions}

\subsection{Freeze-out Parameters} \label{sc.freeze}

\begin{figure}[!htb]
\subfloat[]{
{\includegraphics[scale=0.7]{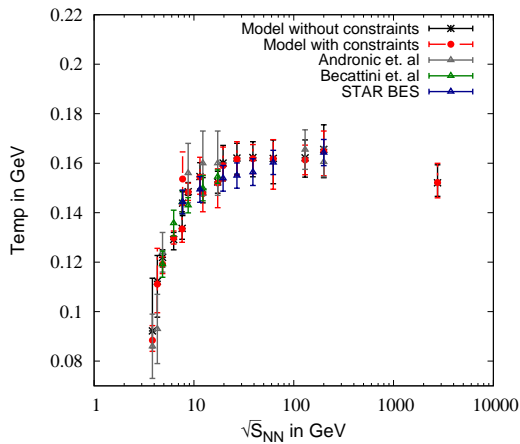}}
\label{fg.Ta}
}\\
\subfloat[]{
{\includegraphics[scale=0.7]{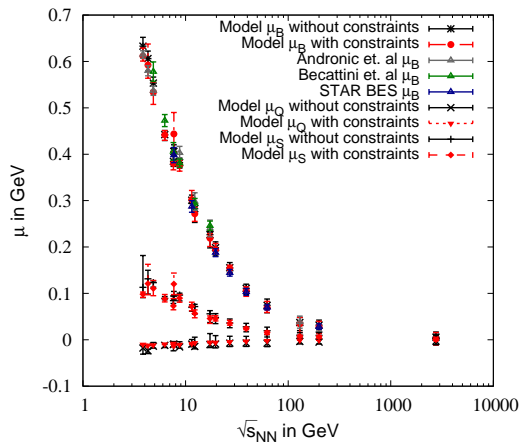}}
\label{fg.Tb}
}\\
\caption{
\label{fg.T}
Variation of $T$, $\mu_B$, $\mu_Q$, $\mu_S$ with $\sqrt{s}$. 
We have compared our freeze-out parameters with Andronic et. al 
~\cite{Andronic:2005yp},  Becattini et. al~\cite{Becattini:2005xt}, STAR 
BES~\cite{Adamczyk:2017iwn}. }
\end{figure}

The variation of the extracted freeze-out parameters with the center of
mass energy $\sqrt{s}$ is presented in Fig.~\ref{fg.T}. The error bars
indicate the variations of the parameters obtained from the different
sets of hadron yield ratios. Results including the constraints on net
charges and those for the unconstrained systems are shown separately.
We find that the extracted parameters in both the schemes are in
reasonable agreement with each other. For comparison, the freeze-out
parameters in some of the recent literature are also plotted. In
general our fitted parameters are commensurate with those reported in
literature with the various forms of the HRG model and also the various
choices of the computation technique. For example
Ref.~\cite{Adamczyk:2017iwn, Andronic:2005yp, Becattini:2003wp,
Becattini:2005xt, Manninen:2008mg, Cleymans:2005xv} have considered a
strangeness suppression factor. For small value of $\mu_Q$, it is
considered to be zero in Ref.~\cite{Adamczyk:2017iwn}. In another work,
the fireball volume was used as a parameter in
Ref.~\cite{Andronic:2005yp, Andronic:2008gu, Andronic:2012ut,
Chatterjee:2015fua}. Interestingly the error bars considered in the
literature, which are obtained by varying the $\chi^2$ by a small
amount, agree well with the systematic variations obtained by us.

In Fig.~\ref{fg.Ta}, the freeze-out temperatures $T$ are shown. The
temperature rises with increasing $\sqrt{s}$ and approaches a
saturation ~\cite{Hagedorn:1965st}, except at the LHC energy where the
temperature is lower~\cite{Abelev:2013vea, ANDRONIC2013535c}. 

The freeze-out chemical potentials $\mu_B$, $\mu_Q$ and $\mu_S$, as
functions of $\sqrt{s}$, are shown in Fig.~\ref{fg.Tb}. The baryon
chemical potential is supposed to be larger for lower $\sqrt{s}$ due to
baryon stopping. This is also why a non-zero charge chemical potential
arise due to the net isospin of the colliding nuclei. On the other hand
a non-zero strangeness chemical potential requires different explanation
as there is no net strangeness in the colliding nuclei. The non-zero
baryon chemical potential induces production of strange baryons. This in
turn requires a strange chemical potential, so that enough strange
anti-particles are produced to ensure the vanishing of net strangeness.
The appearance of a non-zero strangeness chemical potential is perhaps
the strongest signal of the chemical equilibrium of the strongly
interacting system during its evolution~\cite{Witten:1984rs}.

One would expect such a picture to hold naturally in the scheme where
the net charge constraints have been imposed, ensuring the vanishing of
net strangeness. However we find that the strange chemical potential to
follow the same quantitative behavior even when the constraints are
removed. In fact all the freeze-out parameters agree quite well within
the systematics. It appears that the system is so close to chemical
equilibrium that the inclusion or exclusion of external constraints
become irrelevant.

\begin{figure}[!htb]
{\includegraphics[scale=0.7]{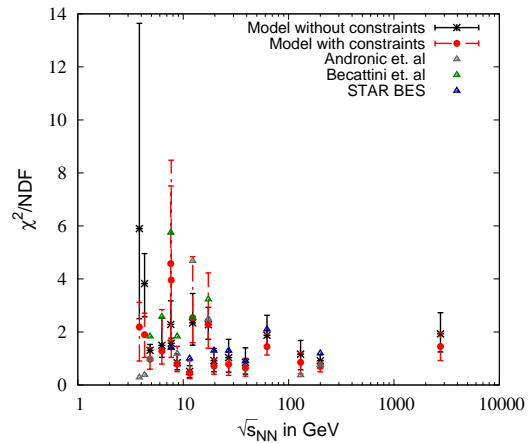}}
\caption{
\label{fg.chisquare}
Variation of $\chi^2$/NDF with $\sqrt{s}$. 
We have compared our $\chi^2$/NDF with 
Andronic et. al ~\cite{Andronic:2005yp},  
Becattini et. al~\cite{Becattini:2005xt}, 
STAR BES~\cite{Adamczyk:2017iwn}}.
\end{figure}

The goodness of the fit can be determined by the reduced $\chi^2$ i.e.
the ratio of $\chi^2$ with the number of degrees of freedom (NDF), which
is shown in Fig.~\ref{fg.chisquare} as a function of $\sqrt{s}$. We show
the comparison between our two schemes of $\chi^2$ minimization and
also plot the reported results available in the literature. We find the
reduced $\chi^2$ in the two schemes to be generally in agreement with
each other. We reiterate that the variations for each scheme are due to
the various choice of independent sets of hadron yield ratios. The
widest variations observed are for the SPS data set, and for the lowest
AGS data set.

As seen from Fig.~\ref{fg.chisquare}, our results are in close proximity
to the results already reported in the literature. It is a general norm
to choose a given set of particle ratios so that the reduced $\chi^2$
lie close to 1. However this is not a necessary condition. The only
requirement is that if we obtain the reduced $\chi^2$ for the variety of
independent sets of hadron yield ratios, they should have a $\chi^2$
distribution with the mean value close to the number of degrees of
freedom. Therefore it should not be considered alarming if the reduced
$\chi^2$ value is far from 1. With the few sets of ratios chosen we find
the mean reduced $\chi^2$ to be close to 1 for most of the $\sqrt{s}$. A
detailed study with a larger number of sets will be reported elsewhere.

\subsection{Hadron Yield Ratios}

\begin{figure*}[!htb]
\centering
\subfloat[]{
{\includegraphics[scale=0.8]{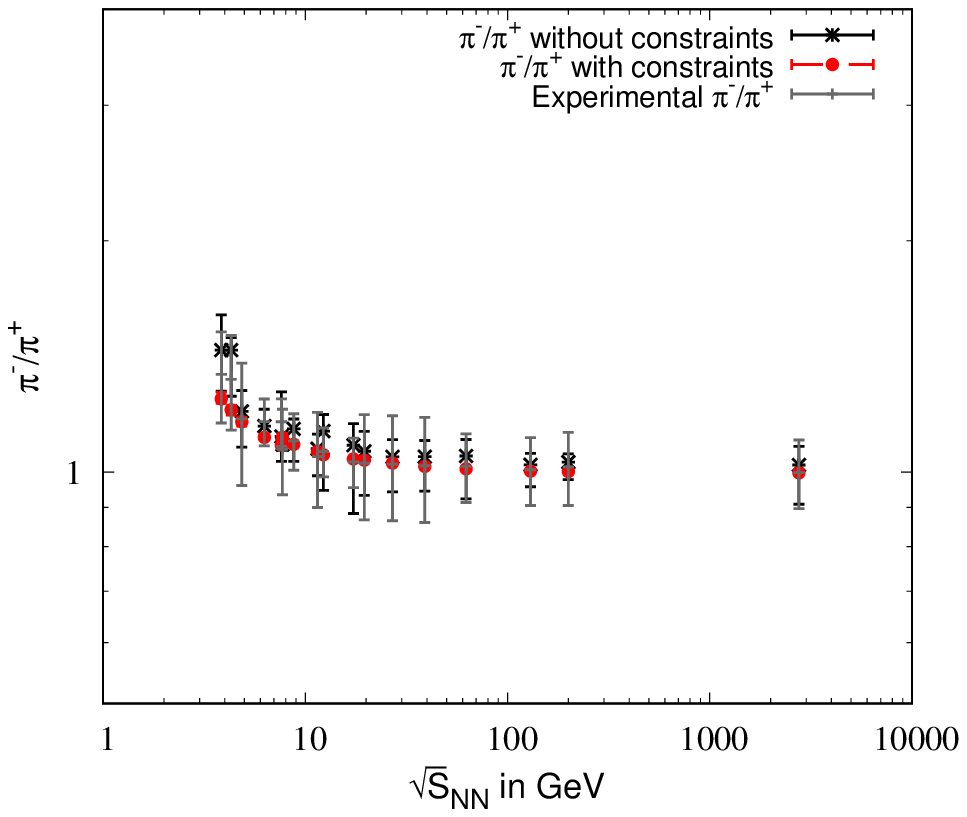}}
\label{fg.pimpip}}
\subfloat[]{
{\includegraphics[scale=0.8]{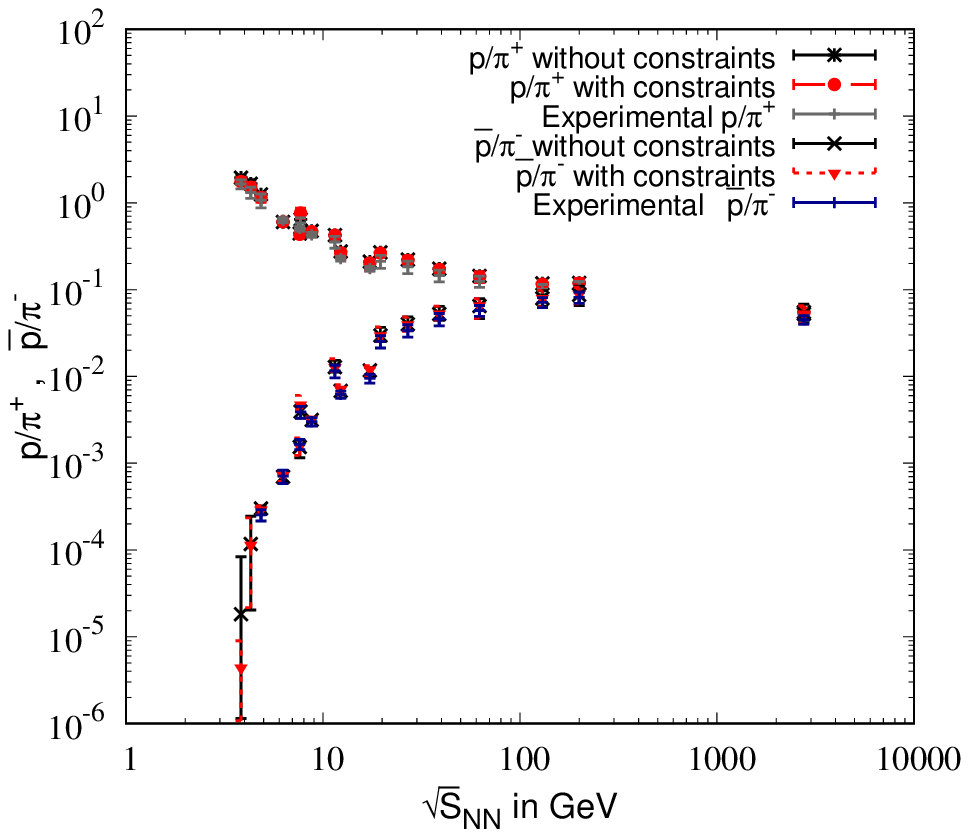}}
\label{fg.ppi}}\\
\subfloat[]{
{\includegraphics[scale=0.8]{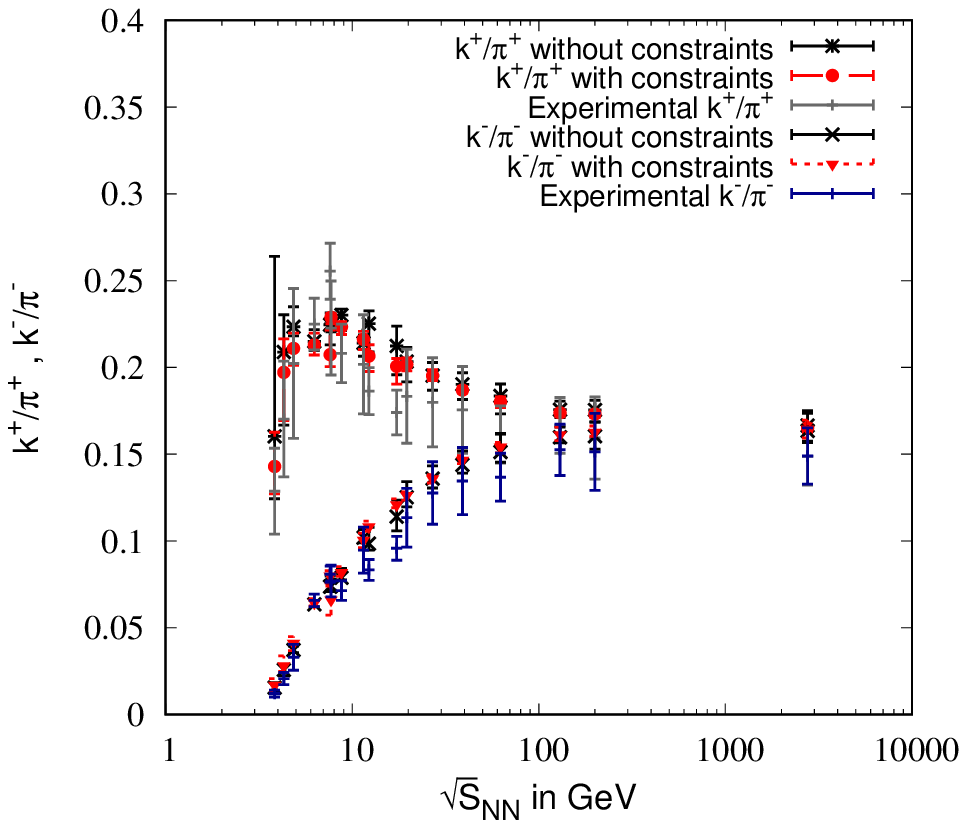}}
\label{fg.kpia}}
\caption{
\label{fg.hratio}
Variation of $\pi^-/\pi^+$, $p/\pi^+$, $\bar{p}/\pi^-$, $k^+/\pi^+$
and $k^-/\pi^-$ with $\sqrt{s}$}
\end{figure*}

\begin{figure*}[!htb]
\centering
\subfloat[]{
{\includegraphics[scale=0.8]{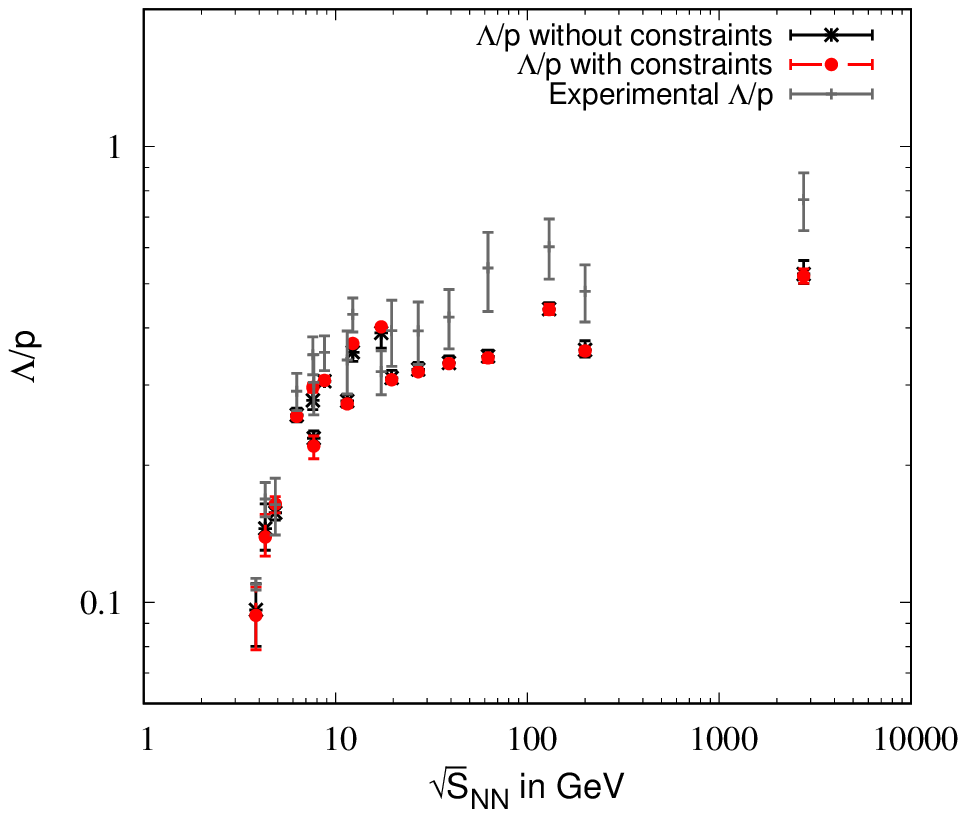}}
\label{fg.lpratio}}
\subfloat[]{
{\includegraphics[scale=0.8]{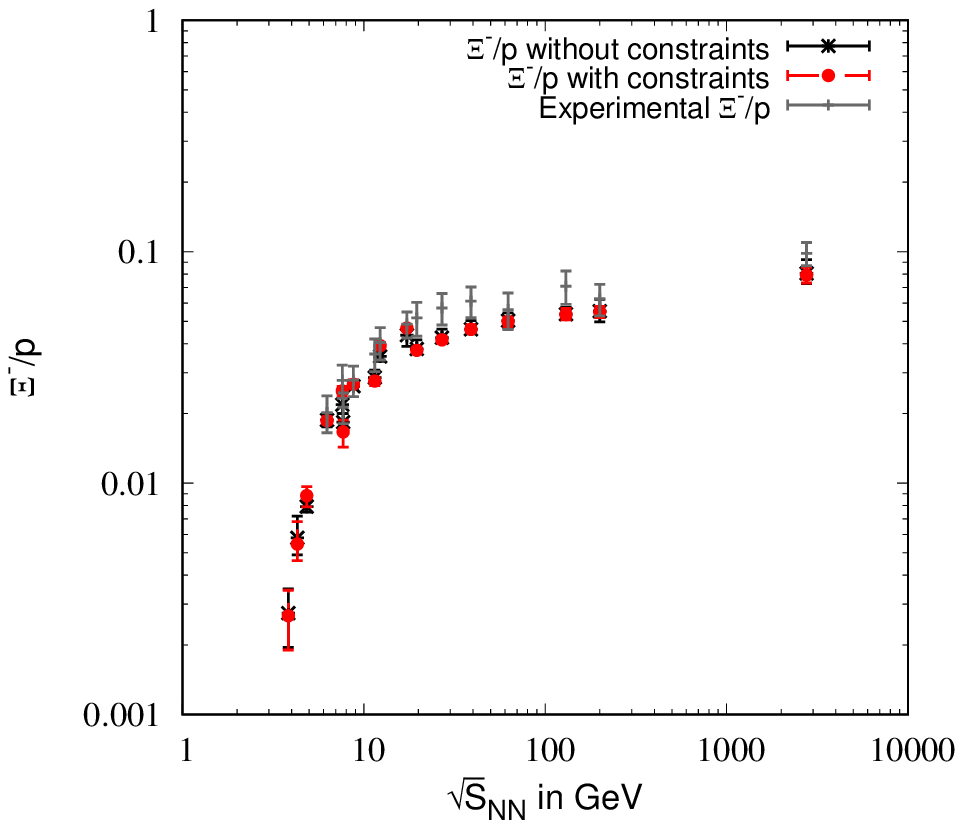}}
\label{fg.xpratio}}
\caption{
\label{fg.hstrange}
Variation of ${\Lambda}/p$ and $\Xi^-/p$ with $\sqrt{s}$}
\end{figure*}

\begin{figure*}[!htb]
\centering
\subfloat[]{
{\includegraphics[scale=0.8]{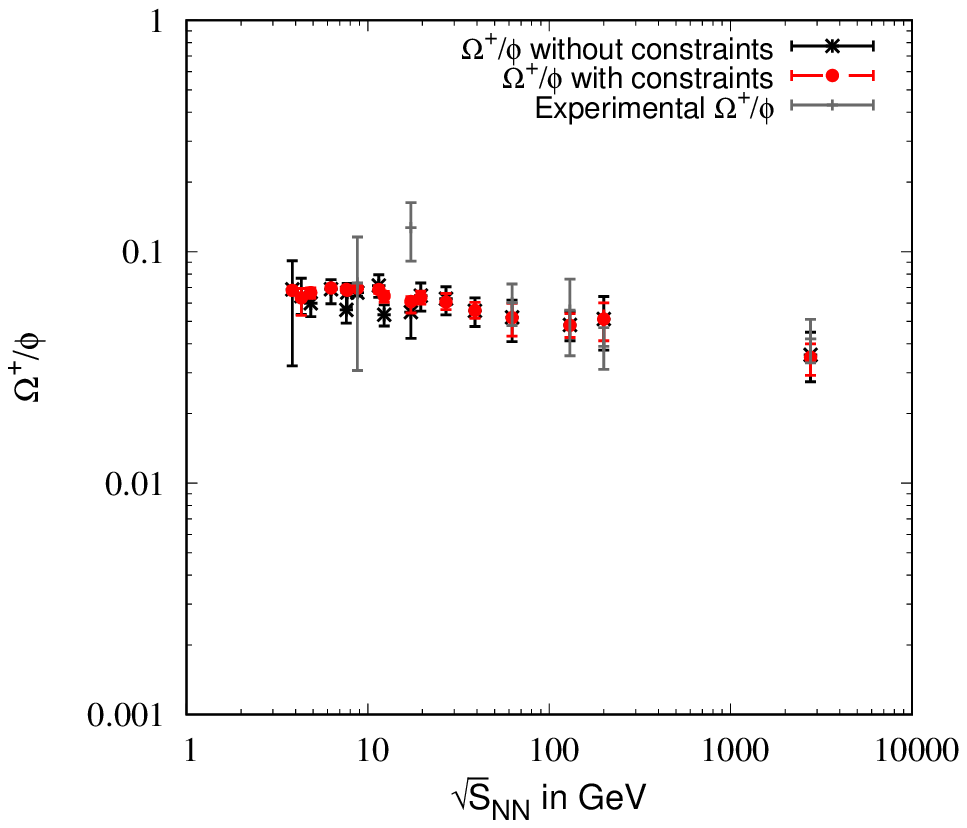}}
\label{fg.omgaphi}}
\subfloat[]{
{\includegraphics[scale=0.8]{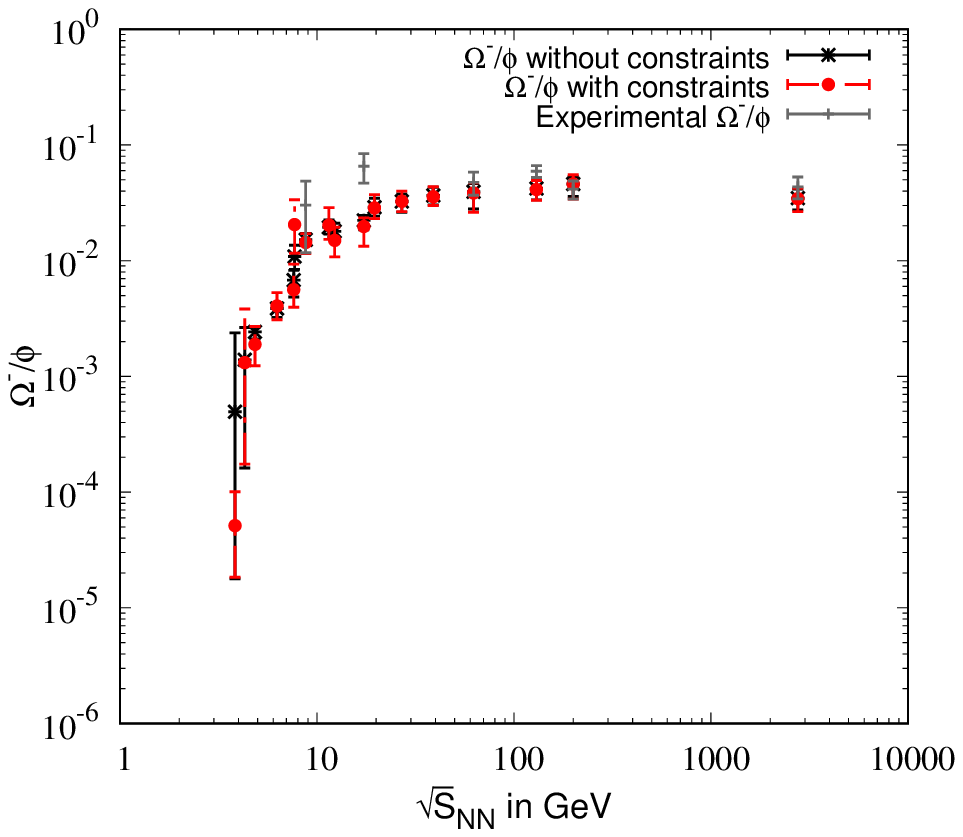}}
\label{fg.omgabarphi}}
\caption{
\label{fg.predratio}
Variation of $\Omega^+/\phi$ and $\Omega^-/\phi$ with
$\sqrt{s}$}
\end{figure*}

With the fitted freeze-out parameters we now obtain the various hadron
yield ratios and compare them with the experimental data. Here we shall
discuss only some of the relevant hadron ratios. The variation of
$\pi^-$ to $\pi^+$ ratio with collisional energy is shown in the
Fig.~\ref{fg.pimpip}. This ratio depends only on $T$ and $\mu_Q$.  The
slight over abundance of neutrons to that of protons in colliding nuclei
leads to an isospin asymmetry yielding more $\pi^{-}$ over $\pi^{+}$.
Thus the experimental $\pi^-$ to $\pi^+$ yield ratio is greater than 1
for lower $\sqrt{s}$, eventually approaching 1 at higher collision
energies. The extracted $\mu_Q$ is therefore slightly negative for small
$\sqrt{s}$ and tends to zero for the higher $\sqrt{s}$. The ratios
obtained from our parametrization for both the schemes follow the same
trend throughout the collisional energy range and reproduced the
experimental data quite well.

In Fig~\ref{fg.ppi}, the extracted ratios $p/\pi^+$ and $\bar{p}/\pi^-$
are similarly found to agree well with the experimental data.  These
ratios are dependent on the interplay between $T$, $\mu_B$ and $\mu_Q$.
At lower $\sqrt{s}$, $p > \bar{p}$ and gradually approaches 1.  On the
other hand the $\pi^-$ is always almost close to $\pi^+$. This leads to
the converging trend of the two shown ratios at higher $\sqrt{s}$.

The same trend is also followed by the $k^+/\pi^+$ and $k^-/\pi^-$
ratios as shown in Fig.~\ref{fg.kpia}. Apart from $T$ and $\mu_Q$, this
ratio depends on $\mu_S$ rather than $\mu_B$. However the behavior of
the $k^+/\pi^+$ ratio is reversed for the lower $\sqrt{s}$, giving rise
to a $'horn'$ like structure near $\sqrt{s} \simeq 10$GeV. This was
originally suggested as an observable for strangeness enhancement, a
signature of onset of deconfinement and QGP
formation~\cite{Gazdzicki:1998vd, Gazdzicki:2003fj,
Bratkovskaya:2004kv,Koch:2005pk}. The explanation of this behavior may
well be beyond the scope of the HRG model, but the model predictions
agree well with the experimental data.

Dependence of $\Lambda/p$ and $\Xi^-/p$ ratios on collisional energy are
shown in the Fig.~\ref{fg.lpratio} and Fig.~\ref{fg.xpratio},
respectively.  The agreement for $\Lambda/p$ with experimental data is
reasonable except for a slight under prediction from model analysis.
Consideration of possible uncertainties in contribution from weak decays
may remove this discrepancy~\cite{Abelev:2008ab}.  However, the model
predictions for $\Xi^-/p$ is found to agree with the experimental data.

Finally in Fig.~\ref{fg.predratio} we plot the variation of some
hadron yield ratios whose experimental data were never used in our
analysis. Specifically we choose the multi-strange hadrons $\Omega$ and
$\phi$. Note that though their yields from the experiments are not used
they appear implicitly when we consider their thermal abundance and
their contribution in the decay chain of the hadrons.  Our predictions
seems to be in excellent agreement with experimental data and the
systematic uncertainties for both the $\chi^2$ schemes are within
control. 

This brings us back to the discussion of the bias induced by the choice
of the the set of yield ratios for obtaining the freeze-out parameters.
As discussed earlier that various authors used {\it suitable} sets of
hadron ratios so that the value of the $\chi^2$ is close to 1. We
discussed above that this is not a necessity. Further, our results show
that the estimated systematic uncertainties due to the choice of
different hadron ratio sets are all within the experimental bounds.
Among the sample ratio sets, while the $\pi^-/\pi^+$ ratio has been used
in all the nine sets considered, the other four ratios have only been
used in one among those sets. Still the predicted values are close to
the experimental data. Obviously this would not have been possible if
the system is very far from the equilibrium description.  However we
should note that only a small subset of all possible sets of independent
ratios were actually used in our study.

In a recent work~\cite{Bhattacharyya:2019wag}, we proposed an alternate
scheme to the $\chi^2$ analysis for studying the freeze-out
characteristics. The basis of that work was to argue that one should use
the quantities associated with various conserved charges of strong
interactions to obtain the freeze-out parameters. But sufficiently close
to equilibrium either of these approaches would produce commensurate
results. Our present results with the considered systematics are found
to be quantitatively in agreement with that alternate prescription.

\subsection{Effects of removing the net charge constraints}

\begin{figure}[!htb]
{\includegraphics[scale=0.7]{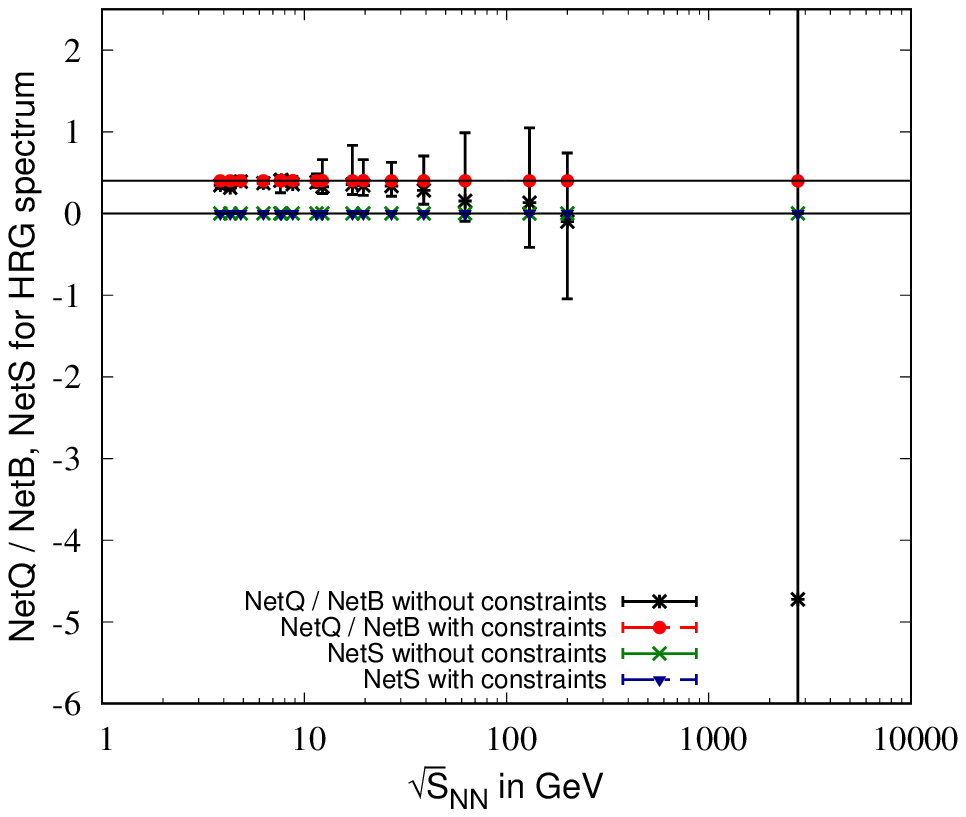}}
\caption{ 
\label{fg.constraints}
Variation of Net$Q$/Net$B$ and Net$S$ with $\sqrt{s}$}
\end{figure}

We finally discuss the predicted behavior of the net charges for the
unconstrained system. In Fig.~\ref{fg.constraints}, we show the
variation of Net$Q$/Net$B$ as well as the Net$S$ as a function of
$\sqrt{s}$. We find that the Net$S$ remains consistent with zero for the
full range of collision energy even without any constraint being put in.
It seems that for chemical equilibration, the system is essentially
guided by the strangeness sector both in terms of the strange chemical
potential as well as the net strangeness. At the same time the
Net$Q$/Net$B$ is also consistent with the expected value of 0.4 up to
$\sqrt{s}=39$ GeV. Thus even without any constraint the system seems to
obey the relations. If the constraints should hold in the central
rapidity region when baryon stopping is significant, then possibly
beyond $\sqrt{s}\sim 40$ GeV, baryon stopping is no longer favorable.

For higher $\sqrt{s}$ the central value of Net$Q$/Net$B$ seems to
decrease, along with a larger systematic uncertainty that increases with
increasing $\sqrt{s}$. The individual Net$Q$ and Net$B$ are
unconstrained inside the central rapidity region so long as these
charges are conserved globally. So there can be possible uncertainties
in their ratios depending on the chosen set of hadron yield ratios.
Also for very large collision energies both of these charges should
approach zero.  This would imply that the ratio could fluctuate
depending on which of the two charges are estimated to vanish with
respect to the other charge. The systematic uncertainty is therefore
found to increase with $\sqrt{s}$ and becomes maximum at the LHC energy.

\section{\label{sec:conclusion} Summary and Conclusion}

Most of the existing literature employ the $\chi^2$ analysis as the
method of choice for extracting the chemical freeze-out parameters from
the available hadron yields. The usual practice is to choose a suitable
set of hadron ratios such that the $\chi^2$ is minimized with respect to
the freeze-out parameters close to the value of $\chi^2/NDF=1$. To
quantify the sensitivity of such a bias, we performed the $\chi^2$
analysis with various sets of hadron yield ratios. We find that the
variation of the extracted chemical freeze-out parameters are very much
under control, as the resulting systematic uncertainties for the hadron
yield ratios lie within close proximity to the experimental bounds.  

Moreover in the standard $\chi^2$ scheme there are two common
assumptions. The first one is that of a fixed net charge to net baryon
number ratio depending on the colliding nuclei, and the other is that of
strangeness neutrality. Although these should indeed be true for the
whole phase space of a given collision event, they may not necessarily
hold for the central rapidity bin. Here we studied a parallel framework
where these conventional constraints are removed. Even in this case the
systematics of the hadron yield ratios seems to completely overlap with the 
system including the constraints.

For the thermodynamic equilibration of the hadrons emerging from the
heavy-ion collision experiments there could be many possibilities. For
example, there could be the equilibration of different species of
hadrons separately, or there could be the equilibration of groups of
certain hadrons. The most interesting possibility is the equilibration of
all the hadrons simultaneously under the umbrella of the conserved
charges of strong interactions. The main intention of our present work
was to find out the extent of such an equilibration achieved in the
heavy-ion collision experiments. If the systematics studied here
resulted in too large variations one would have to conclude that the
holistic equilibration picture does not hold. However our results seem
to indicate that for the full HRG spectrum chemical equilibration in
terms of the conserved charges of strong interactions has been achieved
to a very high degree for a wide range of $\sqrt{s}$ in the heavy-ion
collisions. 

Another outcome of our study is the variation of the Net$Q$/Net$B$ ratio
and net strangeness with $\sqrt{s}$ for the scheme without constraints.
Surprisingly we find the strangeness neutrality to be still valid in
the full range of $\sqrt{s}$. This could only be achieved by an
intricate adjustment of the strangeness chemical potential with respect
to its counterparts for the other conserved charges. This reaffirms that
all the three conserved charges work in unison to bring about the
equilibration of hadronic matter in heavy-ion collisions.

We find the Net$Q$/Net$B$ to remain at its intended value up to $\sqrt{s}
\sim 40$ GeV, and thereafter generally decrease with $\sqrt{s}$ with
growing uncertainties. This could be an indicator of the range of
collision energies beyond which baryon stopping is reduced.

\section{Acknowledgements}
This work is funded by CSIR, UGC and DST of the Government of India.  SB
thanks A. Andronic and J. Stachel for their useful suggestions. 

\bibliography{ref}

\end{document}